\documentclass{frontiersSCNS} 

\usepackage{graphicx}
\usepackage{mathrsfs}
\usepackage{amsmath}
\usepackage{amssymb}
\usepackage{amsfonts}
\usepackage[normalem]{ulem}
\usepackage{color}

\newcommand*{\ket}[1]{\mathopen{\vert}#1\mathclose{\rangle}}

\newcommand*{\bra}[1]{\mathopen{\langle}#1\mathclose{\vert}}

\usepackage{url,lineno}

\copyrightyear{}
\pubyear{}

\def\journal{Physical Chemistry and Chemical Physics}
\def\DOI{}
\def\articleType{Research Article}
\def\keyFont{\fontsize{8}{11}\helveticabold }
\def\firstAuthorLast{Lemeshko} 
\def\Authors{Mikhail Lemeshko\,$^{1,2,*}$}
\def\Address{$^{1}$ITAMP, Harvard-Smithsonian Center for Astrophysics, 60 Garden Street, Cambridge, MA 02138, USA \\
$^{2}$Physics Department, Harvard University, 17 Oxford Street, Cambridge, MA 02138, USA}
\def\corrAuthor{Mikhail Lemeshko}
\def\corrAddress{60 Garden Street, MS-14, Cambridge, MA 02138, USA}
\def\corrEmail{mikhail.lemeshko@gmail.com}

\begin{document}
\onecolumn
\firstpage{1}

\title[Manipulating collisions with controlled dissipation]{Manipulating scattering of ultracold atoms with light-induced dissipation}
\author[\firstAuthorLast ]{\Authors}
\address{}
\correspondance{}
\editor{}
\topic{Chemical Physics with Ultracold Atoms}

\maketitle
\begin{abstract}

\section{}
Recently it has been shown that pairs of atoms can form metastable bonds due to non-conservative forces induced by dissipation [Lemeshko\&Weimer, Nature Comm. \textbf{4}, 2230 (2013)]. Here we study the dynamics of interaction-induced coherent population trapping -- the process responsible for the formation of dissipatively bound molecules. We derive the effective dissipative potentials induced between ultracold atoms by laser light, and study the time evolution of the scattering states. We demonstrate that binding occurs on short timescales of $\sim10~\mu$s, even if the initial kinetic energy of the atoms significantly exceeds the depth of the dissipative potential. Dissipatively-bound molecules with preordained bond lengths and vibrational wavefunctions can be created and detected in current experiments with ultracold atoms.
\tiny
 \keyFont{ \section{Keywords:} ultracold atoms, cold controlled collisions, coherent population trapping, dark state, dissipative preparation of quantum states, dissipatively-bound molecules, non-Hermitian Hamiltonian, Rydberg dressing} 
\end{abstract}

\section{Introduction}

Most realistic systems are ``open'', i.e. coupled to a fluctuating environment, which, for sufficiently strong coupling strengths, is capable of  fundamentally changing the system's properties. In some applications, such as quantum information [\cite{NielsenQIP10}] and coherent spectroscopy [\cite{DemtroderLaserSpec}], the uncontrollable dissipation due to the environment results in decoherence, complicating preparation and read-out of quantum states. In other situations, the environment can lead to novel effects, such as enhanced efficiency of photosynthetic energy transfer in biological systems [\cite{LambertNatPhys12}] and the localization transition in a spin coupled to a bosonic bath [\cite{LeggettRMP87}].

Apart from the fundamental perspective, understanding the effects of environment is crucial for practical applications, since many technologies operate in far-from-equilibrium conditions. In polyatomic systems, usually studied in chemistry and physics, acquiring such an understanding is challenged by the complexity of an underlying Hamiltonian and the uncontrollable nature of dissipation. However, a tremendous recent progress in designing controllable quantum settings paves the way to a detailed understanding of open quantum systems. For example, experimental setups based on ultracold atoms, quantum dots, and superconducting circuits, allow to engineer desired Hamiltonians and control couplings to the environment, thereby getting insight into the microscopic nature of dissipation [\cite{LaddNatPhys10, NatQantSim12, MullerAdv12}]. 
Moreover, the degree of control achieved in such experiments allows to make a step beyond studying the couplings between a system and its environment: recently it has been theoretically predicted that  dissipation can be used as a resource  for  quantum state engineering [\cite{Diehl2008, Verstraete2009, Weimer2010, Diehl2010}]. The method is based upon tuning the the properties of the dissipative bath and system-bath couplings in such a way, that the driven dissipative system evolves towards a desired stationary state. The possibility of using dissipation for quantum state preparation has been recently demonstrated in experiments with cold trapped ions by \cite{Barreiro2011}.

In a recent paper,~\cite{LemWeimDiss} demonstrated that controlled dissipation can be used to create metastable bonds between ultracold atoms. Remarkably, such ``dissipatively-bound molecules'' can be formed even if interparticle interactions are purely repulsive. An extension of this idea to many-particle systems allows to dissipatively prepare crystals of ultracold atoms in free space, i.e.\ without artificially breaking the translational symmetry with an optical lattice or harmonic trap [\cite{OtterLem13}]. In this  contribution, we focus on the effect of light-induced dissipation on the scattering properties of ultracold atoms. Using perturbation theory, we derive the effective dissipative potential curves, and study the time-evolution of the scattering states. We show that by appropriately tuning the couplings of the atoms to the environment one can create dissipatively-bound molecules with desired bond lengths and vibrational wavefunctions.

\section{Material \& Methods}
\begin{figure}[t]
 \centering \includegraphics[width=0.4\linewidth]{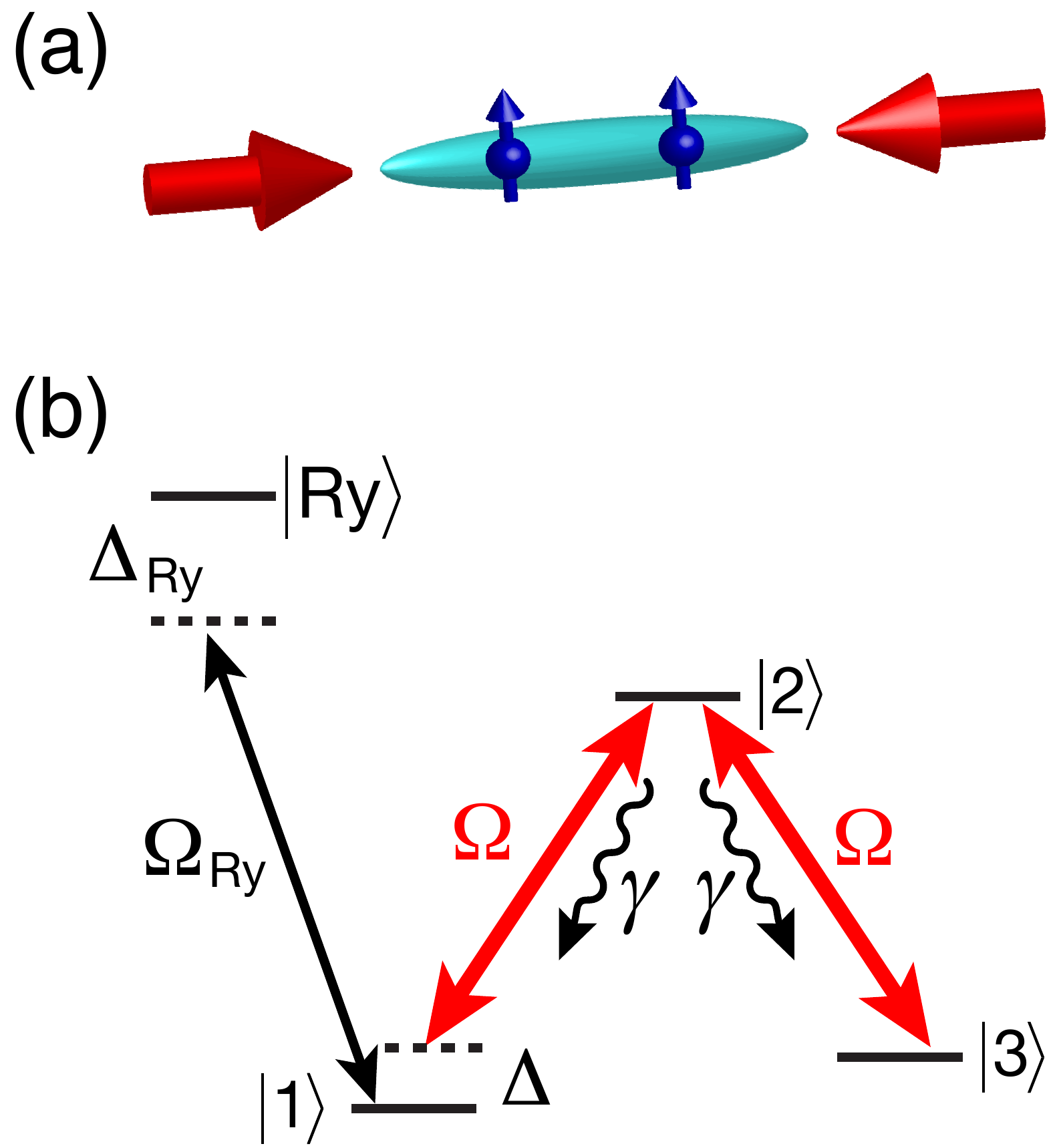}
  \caption{\label{fig:setup} Setup of the system. (a) Ultracold atoms are confined in a one-dimensional optical trap. Two counter-propagating laser beams drive the electronic transitions with a Rabi frequency $\Omega$. (b)~Internal level structure of the atoms. Two components of the ground electronic state,   $\vert 1 \rangle$ and $\vert 3 \rangle$, are coupled to an  electronically excited state,  $\vert 2 \rangle$, spontaneously decaying with a rate $\gamma$. The laser field coupling  states $\vert 1 \rangle$ and $\vert 2 \rangle$ is detuned from the resonance by $\Delta$; state $\vert 1 \rangle$ is provided with an effective dipole moment, $d$, via  far-off-resonant Rydberg dressing, $\Delta_\text{Ry} \gg \Omega_\text{Ry}$. States $\ket{2}$ and $\ket{3}$ are noninteracting.}
\end{figure}

We consider a pair of ultracold atoms whose spatial motion is restricted to one dimension (1D) using an appropriate optical trap, see Fig.~\ref{fig:setup}(a). Each atom possesses the electronic configuration shown in Fig.~\ref{fig:setup}(b). Two fine or hyperfine components of the electronic ground state, $\ket{1}$ and $\ket{3}$, are coupled to an electronically excited state, $\ket{2}$, using two counter-propagating lasers with Rabi frequencies $\Omega$.  In alkali atoms, $\Omega$ corresponds to the laser-cooling transition, $^2S_{1/2} \leftrightarrow {}^2P_{3/2}$. The field coupling states $\ket{2}$ and $\ket{3}$ is on resonance, while the other field is detuned by $\Delta$ from the $\ket{1}-\ket{2}$ transition. For simplicity we assume that $\ket{2}$ decays to both ground states at the same spontaneous emission rate $\gamma$. The atoms are initially prepared in state $\ket{1}$, which is coupled to a highly-excited Rydberg state, $\ket{\text{Ry}}$, possessing a large dipole moment, using a two-photon transition $\Omega_\text{Ry}$ in presence of  a weak external electric field [\cite{Henkel2010,Pupillo2010, Honer2010}]. Dressing of state $\ket{3}$ can be avoided by making use of the dipole selection rules.  If coupling to the Rydberg state  is far-off-resonant, i.e.\ for the detuning $\Delta_\text{Ry} \gg \Omega_\text{Ry}$, one can adiabatically eliminate the state $\ket{\text{Ry}}$ and  assign to  state $\ket{1}$ an effective dipole moment $d$.  As a result, state $\ket{1}$ exhibits a distance-dependent shift induced by the dipole-dipole interaction. On the other hand, states $\vert 2 \rangle$ and $\vert 3 \rangle$ have a zero dipole moment and therefore are non-interacting. We note that a similar setup can be realized based on laser-cooled molecules, that possess nearly closed transitions [\cite{Stuhl2008,ShumanNature10,ManaiPRL12}], in which case the dipole-dipole interactions can be imposed by microwave dressing of rotational levels [\cite{LemeshkoPRL12}].

The $\Lambda$-configuration of Fig.~\ref{fig:setup}(b), formed by two fields $\Omega$, entails a dark state: on resonance, $\Delta=0$, the system is in a stationary state, $\ket{\text{dark}} = (\ket{1} - \ket{3})/\sqrt{2}$, which cannot absorb photons and is therefore decoupled from light. This phenomenon is called coherent population trapping (CPT) [\cite{Gray1978}] and has been used to trap atoms in a particular momentum state below a single photon recoil, so-called velocity selective CPT (VSCPT)~[\cite{Aspect1988, Aspect1989}]. 
In a system of two Rydberg-dressed atoms, the dipole-dipole interaction renders the detuning $\Delta$ dependent on the interatomic distance $r$. This results in interaction-induced CPT: at a particular ``dark distance'', $r_\text{d}$, the interaction shifts level $\ket{1}$ to resonance, effectively decoupling the atomic pair from photon absorption-emission. The resulting metastable state corresponds to a dissipation-induced interatomic bond, recently described by \cite{LemWeimDiss}. In this section we derive the effective potentials corresponding to the non-conservative forces acting between ultracold atoms, which underly the formation of the dissipation-induced bonds.

A pair of ultracold atoms described above represents an open quantum system with the electromagnetic field acting as a reservoir. The system's dynamics is given by the quantum master equation for the density operator  $\rho$ [\cite{Breuer2002}]:
\begin{equation}
\label{eq:MasterEq}
	\frac{d \rho}{d t} = - i / \hbar \left[H, \rho \right] +
        \sum_{n} \gamma_{n} \left(c_{n}^{\phantom{\dagger}} \rho c_{n}^\dagger -
        \frac{1}{2} \{ c_{n}^\dagger c_{n}^{\phantom{\dagger}}, \rho \} \right),
\end{equation}
with $\hbar$ Planck's constant. The coherent part of the dynamics is contained in the Hamiltonian accounting for the motion of the atoms, their interaction with the laser fields, and the dipole-dipole interactions,
\begin{multline}
\label{Hamil}
	H = \sum_{k, i} \Biggl [ \frac{\hbar^2 k^2}{2 m} \vert k \rangle \langle k\vert_i - \frac{\Omega}{2} \left(\vert 1, k+\Delta k \rangle \langle 2, k \vert_i + \text{h.c.} \right) \\
	- \frac{\Omega}{2} \left(\vert 3, k-\Delta k \rangle \langle 2, k \vert_i  + \text{h.c.} \right)  - \Delta \vert 1, k \rangle \langle 1, k \vert_i  \Biggr] \\ 
	+ \sum_{k, k',q} \tilde{V}_\text{dd} (q) \vert 1, k-q\rangle_1 \vert 1, k'+q \rangle_2 \langle 1, k\vert_1\langle 1, k' \vert_2.
\end{multline}
Here, $i = 1,2$ and $k$ label the atoms and their corresponding momentum states, and $\tilde{V}_\text{dd} (q)$ is the Fourier transform of the dipole-dipole interaction potential. The dissipative part of Eq.~(\ref{eq:MasterEq}) contains the rates $\gamma_n = \gamma$ and jump operators $c_n =  \sum_k \vert k+\Delta k_n, j_n \rangle \langle 2, k|_{i_n}$ in the Lindblad form, responsible for the decay of each atom from state $\vert 2 \rangle$. The index $i_n=1,2$ runs over the two atoms, while $j_n=1,3$ accounts for the two final states, and $\Delta k_n$ contains all possible values of the emitted photon's wave vector [\cite{DalibardPRL92, MolmerJOSAB93}].

In the regime of weak dissipation, $\Omega^2/\gamma^2 \ll 1$, one can neglect the quantum jumps, i.e.\ the $c_{n}^{\phantom{\dagger}} \rho c_{n}^\dagger$ term of Eq.~(\ref{eq:MasterEq}). As a result, the dynamics of the system is described by an effective non-Hermitian Hamiltonian,
$H_\text{eff} = H - i V_\text{d}$, containing a dissipative potential:
\begin{equation}
\label{Vd}
V_\text{d} = \hbar \sum\limits_n \frac{\gamma_n}{2} c_n^\dagger c_n^{\phantom{\dagger}}.
\end{equation}

In this work we focus on ultracold atoms at sub-Doppler temperatures of $ \sim 1-10$ microKelvin. In particular, the kinetic energy is considered to be small compared to the dipolar interaction between the atoms, which in turn is small compared to the  laser Rabi frequency and the spontaneous decay rate, i.e., $(\hbar k)^2/2m \ll \Delta, \delta(r) \ll \gamma,\Omega$. 
As a first step, we derive the effective interatomic potentials in the limit of zero kinetic energy, i.e.\ the center-of-mass motion is considered to be cold enough in order to neglect the corresponding Doppler shifts; in this case the resulting effective potentials are independent on the relative momentum.

In a two-atom system, the fields $\Omega$ connect only the states symmetric against the particle exchange, therefore we reduce the manifold of relevant states to the following 6 levels: $\{ \ket{11}; \left(\ket{12} + \ket{21} \right)/\sqrt{2}; \left(\ket{13} + \ket{31} \right)/\sqrt{2}; \ket{22}; \left(\ket{23} + \ket{32} \right)/\sqrt{2}; \ket{33} \}$.  In this basis, the interaction part of the two-atom Hamiltonian reads:

\begin{equation}
\label{Hamil2at}
	H_\text{int} = \left( \begin{array}{c c c c c c}
	 V(r) - 2\Delta & \frac{\Omega}{\sqrt{2}} & 0 & 0 & 0 & 0\\[10pt]
	 \frac{\Omega}{\sqrt{2}} & - \frac{i\gamma}{2} - \Delta & \frac{\Omega}{2} &  \frac{\Omega}{\sqrt{2}}  & 0 & 0\\[10pt]
	 0 & \frac{\Omega}{2} & -\Delta & 0 & \frac{\Omega}{2} & 0 \\[10pt]
	 0 & \frac{\Omega}{\sqrt{2}} & 0 & -i\gamma & \frac{\Omega}{\sqrt{2}} & 0\\[10pt]
	 0 & 0& \frac{\Omega}{2} & \frac{\Omega}{\sqrt{2}} &  - \frac{i\gamma}{2} & \frac{\Omega}{\sqrt{2}} \\[10pt]
	 0 & 0 & 0 & 0& \frac{\Omega}{\sqrt{2}} &0
	 \end{array}   \right)
\end{equation}
where $V(r)$ gives the interaction between the atoms in state $\ket{11}$. Treating $V(r)$ and $\Delta$ as perturbations, one can obtain the ground state, $\ket{\psi}$, of the Hamiltonian (\ref{Hamil2at}), which corresponds to the density operator $\rho = \sum_{i,j} \rho_{ij} \ket{\psi} \bra{\psi}$.  The complex dissipative potential $V_\text{d} (r)$ can be derived as the total probability of photon absorption, i.e.\ as a sum of those density matrix elements $\rho_{ij}$ that correspond to the transitions $\ket{1} - \ket{2}$ and $\ket{3} - \ket{2}$ in each atom, multiplied by the photon scattering rate, $\Omega^2/\gamma$. The resulting expression for  $V_\text{d} (r)$ reads:
\begin{equation}
\label{Vd}
	V_\text{d} (r) = A - B V(r) + C V^2(r),
\end{equation}
where $A = \sqrt{2} \Delta^2 / (2\Omega)$, $B = \sqrt{2} \Delta/\Omega$, and $C = \left[ (4+9\sqrt{2})/(64\Omega) + (8+6\sqrt{2})\Omega/(64 \gamma^2) \right]$.

\section{Results}

\begin{figure}[t]
 \centering \includegraphics[width=0.3\linewidth]{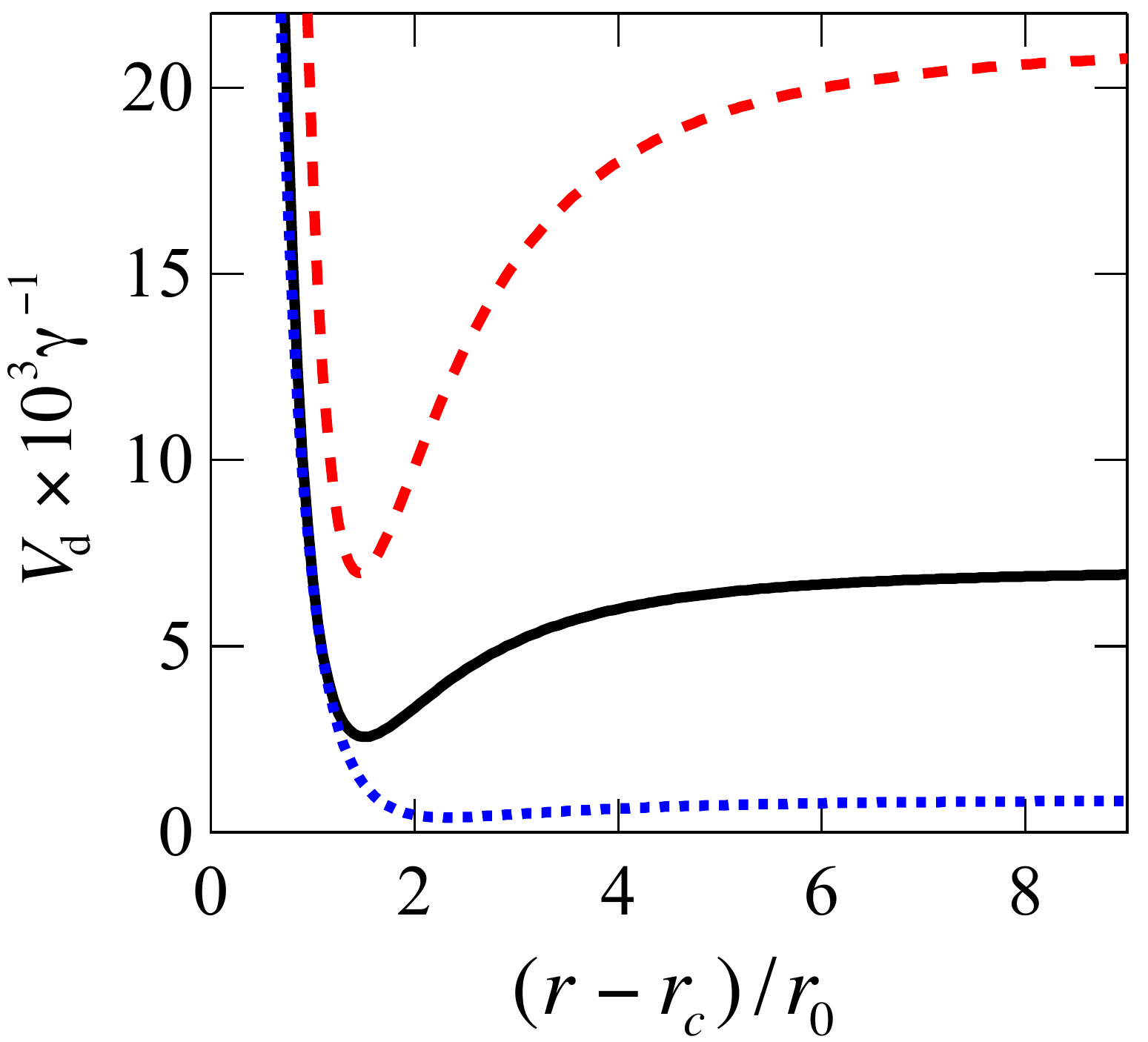}
  \caption{\label{fig:pots} Examples of the dissipative potentials, Eq.~(\ref{VdExpl}), for different values of parameters. Black solid line: $\Omega=\gamma/4$, $\Delta =\gamma/20$ (these values correspond to the results of Figs. \ref{fig:evolution} and \ref{fig:enrad}); blue dotted line: $\Omega=\gamma/2$, $\Delta =\gamma/40$; red dashed line: $\Omega=\gamma/12$, $\Delta =\gamma/20$. The cutoff radius $r_\text{c} = r_0$. The steady-state probability distributions corresponding to these potentials are shown in Fig.~\ref{fig:steady}.}
\end{figure}

In what follows, we assume the dipole-dipole interaction between the atoms, $V(r) = d^2/(4 \pi \epsilon_0 r^3)$, where $d$ is the effective dipole moment of state $\ket{1}$ and $\epsilon_0$ is the vacuum permittivity. Using $\gamma$ as a unit of energy and $r_0 = d^{2/3}/(4 \pi \epsilon_0 \gamma)^{1/3}$ as a unit of distance, we obtain the following shape of the imaginary potential:
\begin{equation}
\label{VdExpl}
	V_\text{d} (r) = C_0 - \frac{C_3}{r^3} + \frac{C_6}{r^6},
\end{equation}
Here $C_0 = \left(\frac{\Delta}{\gamma} \right)^2/\left (\sqrt{2} \frac{\Omega}{\gamma} \right)$, $C_3 =  \left(\frac{\Delta}{\gamma} \right)/\left (\sqrt{2} \frac{\Omega}{\gamma} \right)$, and $C_6 = \left[ (4+9\sqrt{2}) + (8+6\sqrt{2})  \left(\frac{\Omega}{\gamma} \right)^2 \right]/ \left(64 \frac{\Omega}{\gamma}\right)$. The dissipative potential (\ref{VdExpl}) possesses a minimum at the so-called ``dark distance'', $r_\text{d}$, where the photon scattering rate is significantly reduced,
\begin{equation}
\label{rd}
  r_\text{d} =  \left(\frac{a_1+a_2 (\Omega/\gamma)^2}{\Delta/\gamma} \right)^{1/3},
\end{equation}
with  $a_1 = (9+2\sqrt{2})/16$, and $a_2 = (3+2\sqrt{2})/8$. 
The value of $r_\text{d}$, as well as the depth of the potential well, $D=C_6/r_\text{d}^6$, can be tuned by changing the Rabi-frequency and the detuning of the laser fields. Figure~\ref{fig:pots} exemplifies the dissipative potentials for different values of  $\Omega$ and $\Delta$. 

The dynamics of the two-atom system is given by the Schr\"odingier equation in the center-of-mass frame:
\begin{equation}
\label{SchrEq}
i \frac{\partial }{\partial t} \ket{\psi} = -\left[ \alpha \nabla^2 + i V_\text{d} (r) \right ]  \ket{\psi}
\end{equation}
Here the kinetic energy scales with the parameter $\alpha = \hbar^2/(2 m r_0^2 \gamma)$, with $m$ the reduced mass of the atomic pair. Note that $V_\text{d} (r)$ occurs in Eq.~(\ref{SchrEq}) with a minus sign, i.e.\ the distance $r_\text{d}$ corresponds to a reduced absorption of particles.

In order to study the time evolution of the scattering states,  we  solve Eq.~(\ref{SchrEq}) numerically for different initial conditions.
We exemplify the laser driving with the Rabi frequency $\Omega = \gamma/4$ and detuning $\Delta = \gamma/20$, which corresponds to $V_\text{d}(r)$ shown in Fig.~\ref{fig:pots} by the black solid line; this results in the dark distance $r_\text{d} = 2.5 r_0$.  In order to simulate an experiment with a fixed particle density, we consider two particles confined in a box of length $L=14.5~r_0$. We set the kinetic energy parameter to $\alpha = 4\cdot 10^{-4}$, and use a short-range cutoff, i.e. an impenetrable wall condition, for $r<r_\text{c} = r_0$. These values of parameters can be realized, e.g.\, with ultracold cesium atoms. In this case, two hyperfine components of the  ground electronic  state, $6^2S_{1/2}$, are chosen as states $\ket{1}$ and $\ket{3}$; state $\ket{1}$ is provided with an effective dipole moment $d=15$~D due to Rydberg dressing in an external electric field. The laser field $\Omega$ drives the $6^2S_{1/2} \leftrightarrow 6^2P_{3/2}$ transition whose linewidth is $\gamma = 2\pi \times 5.2$~MHz. This results in $r_0 = 186$~nm, which corresponds to the three-dimensional atomic density of $4\cdot10^{11}$~cm$^{-3}$. In our calculation, the spatial grid is chosen such that the maximal value of the relative momentum $k_\text{max} = 18.5/r_0$, which in the case of Cs corresponds to$~\sim14$ atomic recoil momenta.

\begin{figure}[t]
\hspace{-0.6cm} \includegraphics[width=1.03\linewidth]{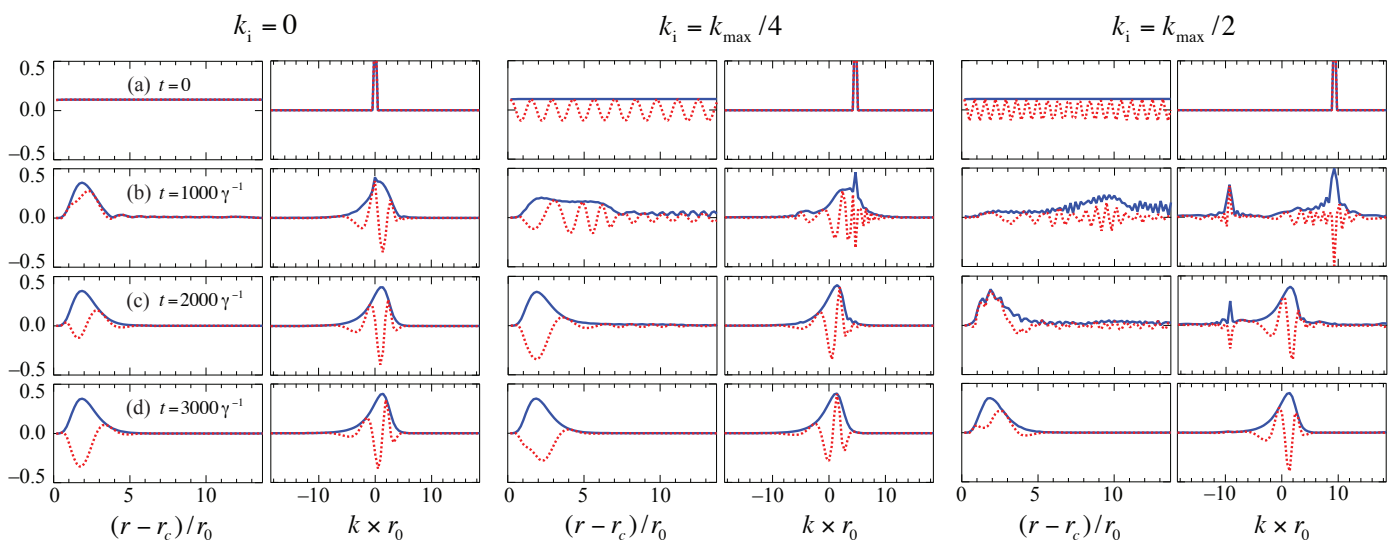}
  \caption{\label{fig:evolution}  Absolute values (blue solid lines) and real parts (red dotted lines) of the scattering wavefunctions at times (a) $t=0$; (b) $t=1000\gamma^{-1}$; (c) $t=2000\gamma^{-1}$; and (d) $t=3000\gamma^{-1}$. Pairs of columns show the wavefunctions in the position and momentum representation for different initial relative momenta: $k_\text{i} = 0$ (left); $k_\text{i} = k_\text{max}/4$ (middle); and $k_\text{i} = k_\text{max}/2$ (right), with $k_\text{max} = 18.5/r_0$. The cutoff radius $r_\text{c} = r_0$.}
\end{figure}

Figure~\ref{fig:evolution} shows the time evolution of the scattering states starting from different initial relative momenta, $k_\text{i}$. Three pairs of columns show the cases of $k_\text{i}=0$ (left panels),  $k_\text{i} = k_\text{max}/4$ (middle panels), and $k_\text{i} = k_\text{max}/2$ (right panels).  Within each pair, the left column shows the wave function in the position representation, $\psi(r,t)$, while the right column shows its Fourier transform, giving the relative momentum distribution of the scattering state, $\psi(k,t)$. 
One can see that in the long-time limit, panels (d), the driven-dissipative dynamics steers the pair of atoms towards the same steady state, independent of the initial conditions. The interplay between the kinetic energy term and the dissipative potential of Eq.~(\ref{SchrEq}) results in a distribution of the relative distances around $r_\text{d}$, and the relative momentum distribution peaked in the vicinity of $k=0$. As a result, the dissipation-induced bond is formed. Similarly to conventional molecules bound by conservative forces [\cite{Herzberg}], the distance and momentum distributions are asymmetric, which arises due to the anharmonicity of the dissipative potential, cf. Fig.~\ref{fig:pots}.

Note that the presented cases of $k_\text{i} = k_\text{max}/4$ and $k_\text{i} = k_\text{max}/2$ correspond to the initial kinetic energies of $9\times 10^{-3} \gamma$ and $34\times 10^{-3} \gamma$ respectively, which significantly exceeds the depth of the dissipative potential well, cf. Fig.~\ref{fig:pots}. Interestingly, even in this case the formation of the dissipative bond occurs at a timescale comparable to the case of $k_\text{i}=0$. For a pair of Cs atoms, the unit of time $\gamma^{-1} \approx 30$~ns, i.e.\ the timescales of the bond formation are on the order of $10-100~\mu$s. Since the initial atomic wavefunctions of Fig.~\ref{fig:evolution}(a) are completely delocalized in space, the bonding timescales are a few times longer compared to the ones obtained in [\cite{LemWeimDiss}].

\begin{figure}[t]
\centering\includegraphics[width=0.3\linewidth]{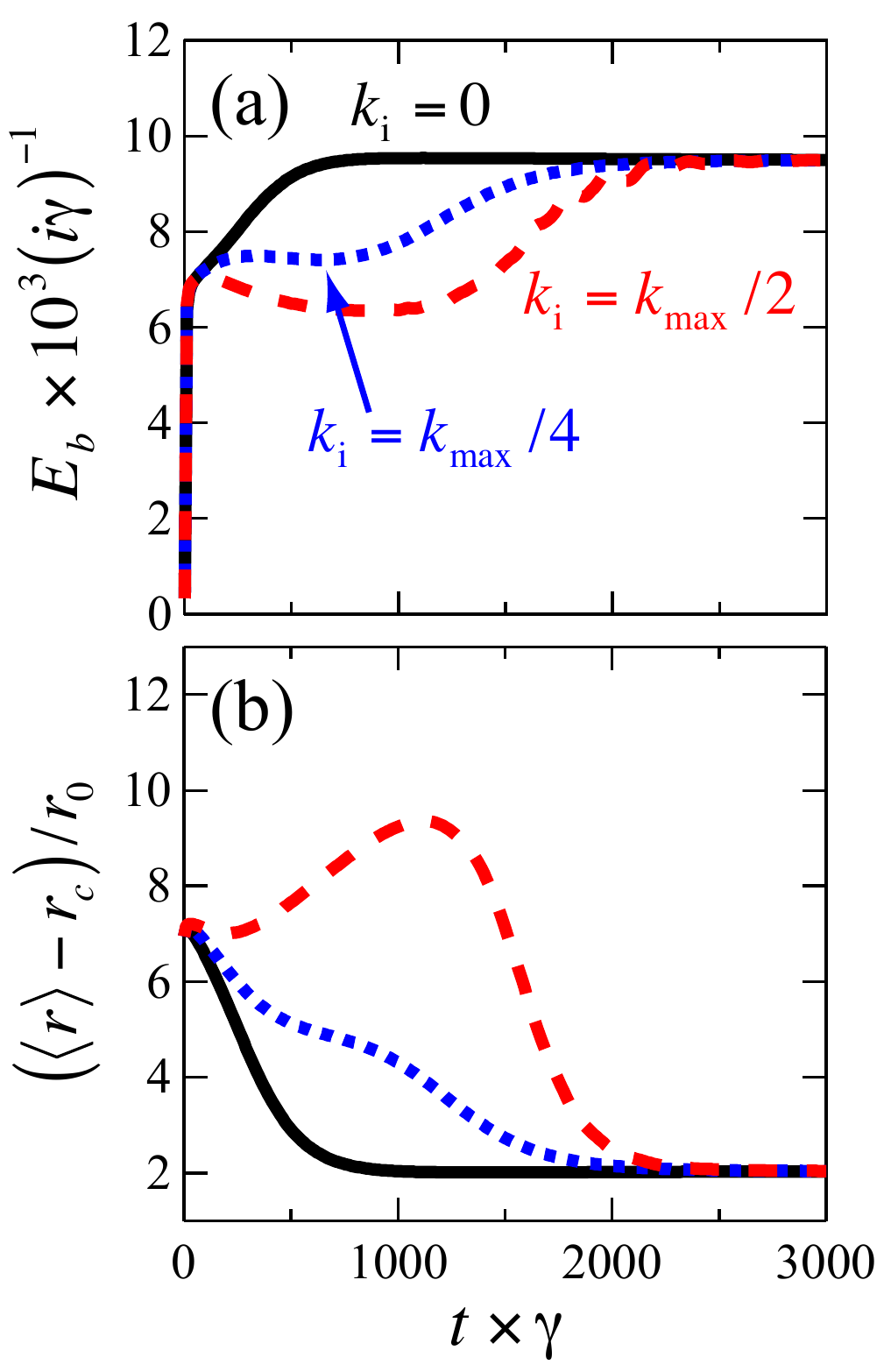}
  \caption{\label{fig:enrad}  (a) Time-dependence of the imaginary binding energy  of the molecules, Eq.~(\ref{Ebind}). (b) Time-dependence of the mean interatomic distance, Eq.~(\ref{Rmean}). Results for different initial relative momenta are shown: $k_\text{i}=0$ (black solid line), $k_\text{i} = k_\text{max}/4$ (blue dotted line), and $k_\text{i} = k_\text{max}/2$  (red dashed line), with $k_\text{max} = 18.5/r_0$. The cutoff radius $r_\text{c} = r_0$.}
\end{figure}

The dynamics of the bond formation can be characterized by the imaginary binding energy. It is defined as the difference between the dissipation rate at $t=0$, corresponding to atoms completely delocalized in space, and the dissipation rate at a given time $t$, when the molecules are formed:
\begin{equation}
\label{Ebind}
	E_\text{b}(t) = i \int  V_\text{d} (r) \left( \vert \psi(r,0)\vert^2 -  \vert \psi(r,t)\vert^2 \right) dr
\end{equation}
Figure~\ref{fig:enrad}(a) shows the time evolution of $E_\text{b}$ starting from different initial conditions. At small times $t$ the binding energy grows rapidly, due to strong dissipation at small interatomic distances that quickly pushes the population towards larger $r$. In the long-time limit, $E_\text{b}$ approaches the value of $9.5 \times 10^{-3} i \gamma$, independently of the initial relative momentum. Note that while the qualitative behavior of $E_\text{b}(t)$ does not depend on the details of the interatomic potential, the lower limit of the integral in Eq.~(\ref{Ebind}) is set by the short-range cutoff radius $r_\text{c}$. The exact numerical value of $E_\text{b}$ therefore depends on $r_\text{c}$.

The length of the dissipative bond is characterized by the mean interatomic distance,
\begin{equation}
\label{Rmean}
	\langle r \rangle (t)  = \int  \vert \psi(r, t)\vert^2 r dr,
\end{equation}
whose time-evolution is shown in Fig.~\ref{fig:enrad} (b). In the long-time limit $\langle r \rangle$ approaches the value of $\langle r \rangle \approx 3 r_0 \approx 1.2 r_\text{d}$.


\begin{figure}[t]
 \centering \includegraphics[width=0.4\linewidth]{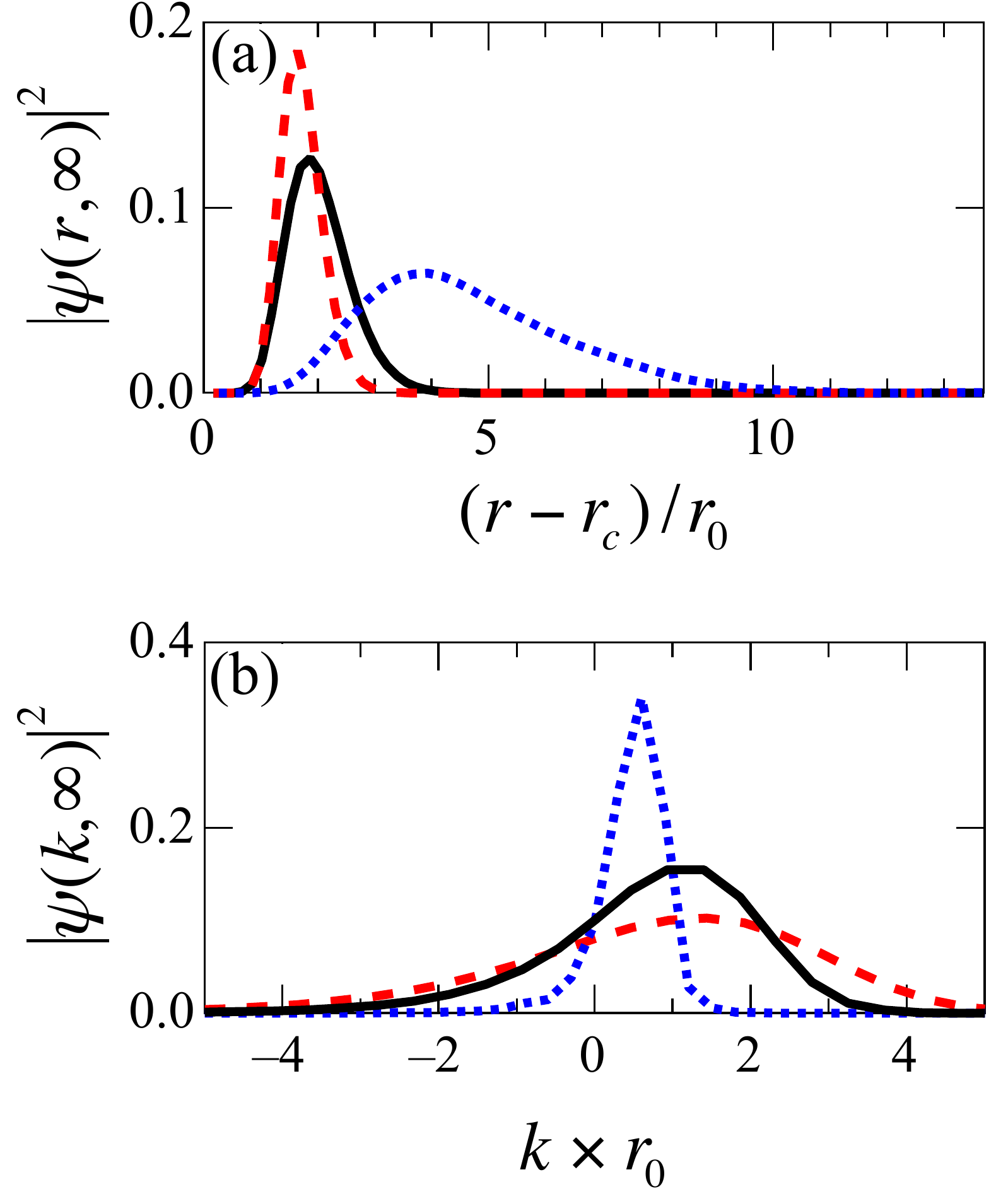}
  \caption{\label{fig:steady} Steady states corresponding to the potentials of Fig.~\ref{fig:pots}. (a) Probability distribution of relative distances; (b) corresponding relative momenta distributions. Black solid line: $\Omega=\gamma/4$, $\Delta =\gamma/20$; blue dotted line: $\Omega=\gamma/2$, $\Delta =\gamma/40$; red dashed line: $\Omega=\gamma/12$, $\Delta =\gamma/20$. The cutoff radius $r_\text{c} = r_0$.}
\end{figure}

\section{Discussion}

In this work we studied collisions of ultracold atoms in presence of dissipation due to near-resonant scattering of laser photons. The laser configuration is chosen such, that dissipation is significantly reduced at some preordained interatomic distance $r_\text{d}$, due to the interaction-induced coherent population trapping taking place. Working in the regime of small kinetic energy, we derived the effective, purely imaginary, interatomic potentials featuring minima at $r=r_\text{d}$, whose shape can be tuned by changing the laser Rabi frequency and detuning. 

Starting from the states with a particular value of the relative momentum, we studied the time evolution of the scattering wavefunctions.  It was shown that, independently of the initial conditions, the driven-dissipative dynamics results in the steady state corresponding to a dissipatively-bound atomic pair, reached at the short timescales of $\sim10-100~\mu$s. Interestingly, the bound states are formed even if the initial kinetic energy significantly exceeds the depth of the dissipative potential well. The dynamics of the dissipation-induced association was characterized by the time-dependent relative distance and momentum distributions, bond lengths, and imaginary binding energies.

The spectroscopic parameters of the dissipatively-bound molecules can be altered by tuning the laser parameters: the bond length scales with the effective dipole moment $d$ due to Rydberg dressing, and with the detuning $\Delta$; the binding energy (and therefore the shape of the vibrational wavefunction) depends on the photon scattering rate proportional to $\Omega^2/\gamma$. Fig.~\ref{fig:steady}(a) shows the vibrational probability distributions for molecules bound by the potentials shown in Fig.~\ref{fig:pots}; the panel (b) shows the corresponding momenta distributions. One can see that appropriately choosing the laser frequency and intensity makes it possible to prepare molecules with desired vibrational wavefunctions.

The relative distance distributions shown in Fig.~\ref{fig:evolution} and \ref{fig:steady}(a) are proportional to the pair-correlation function, $g^{(2)}(r)$, that can be directly measured in experiment using a number of techniques such as noise correlation spectroscopy [\cite{Altman2004,Folling2005}] or Bragg scattering [\cite{Stamper-Kurn1999}]. The dissipative bond manifests itself as a peak emerging in $g^{(2)}(r)$ during the time evolution.
It is worth noting that dissipative binding occurs during the incoherent evolution of the scattering states, therefore its observation does not require long coherence times needed to observe the effects of the interactions on the coherent evolution of Rydberg-dressed atoms. The coherence times longer than the binding timescales of $\sim0.1$~ms are achievable in current experiments [\cite{Low2012, Schempp2010, Pritchard2010, Nipper2012, Dudin2012, Peyronel2012,  SchaussNat12}].

The goal of this work was to develop a simple model allowing to understand the main features of dissipation-assisted scattering and  the dynamics of interaction-induced coherent population trapping. The model is based on a few approximations whose limitations are worth discussing here. First, within the wavefunction approach that we employed, the system's state was assumed to be pure, as opposed to a mixed state resulting from the solution of the full master equation, eq.~(\ref{eq:MasterEq}). Furthermore, the applied theory did not include quantum jumps that might quantitatively alter the  relative momentum distribution in the steady-state, as well as the evolution times at which the steady state is reached. Finally, the internal and external degrees of freedom were decoupled from each other by introducing an effective interatomic potential, therefore the model does not provide information about the final population of the ground states $\ket{1}$ and $\ket{3}$. However, even with these approximations in place, the model is capable of capturing the  physics of the system, as it is confirmed by a good agreement with the results of Ref.  [\cite{LemWeimDiss}].

Finally, while in this work we focused on the realization based on Rydberg-dressed   atoms [\cite{Henkel2010,Pupillo2010, Honer2010}], similar ideas can be applied to laser-cooled polar molecules [\cite{Stuhl2008,ShumanNature10,ManaiPRL12}]. Extensions to other types of interparticle interactions, such as magnetic dipole-dipole [\cite{LuPRL11}] and electric quadrupole-quadrupole [\cite{BhongalePRL13}] ones, also seem possible.

\section*{Disclosure/Conflict-of-Interest Statement}
The authors declare that the research was conducted in the absence of any commercial or financial relationships that could be construed as a potential conflict of interest.

\section*{Acknowledgement}
I am grateful to Hendrik Weimer and Johannes Otterbach for fruitful discussions. 

\paragraph{Funding\textcolon} The work was supported by the NSF through a grant for the Institute for Theoretical Atomic, Molecular, and Optical Physics at Harvard University and Smithsonian Astrophysical Observatory.


\bibliographystyle{frontiersinSCNS_ENG} 
 \bibliography{Diss_scat}

\end{document}